# Monopole polarization of $C_{60}$ fullerene shell by an electric charge in its center

M. Ya. Amusia[1,2] and A. S. Baltenkov[3]

[1] Racah Institute of Physics, the Hebrew University, Jerusalem, 91904 Israel
[2] Ioffe Physical-Technical Institute, St. Petersburg, 194021 Russia
[3] Arifov Institute of Ion-Plasma and Laser Technologies,
Tashkent, 100125 Uzbekistan

**Abstract**
The spatial distributions of electric charges forming the $C_{60}$ shell have been analyzed with the Poisson equation. It has been shown that the modification of formulas for the rectangular-potential-well as a model potential of the $C_{60}$ shell by means of addition to them the Coulomb-potential-like terms does not describe the monopole polarization of the shell by the electric charge $Z$ in the center of the shell. The phenomenological potentials simulating the $C_{60}$ shell potential should belong to a family of potentials with a non-flat bottom. In the paper some model potentials have been proposed and discussed. It has been shown that by varying the semithiknesses of these potential wells we could describe the various degrees of the monopole polarization of the $C_{60}$ shell that due to shifting of collectivized electrons of the shell relative to the rigid positive carcass of $C_{60}$ fullerene. Constructed here, the model potentials for the $C_{60}$ shell can be used to describe the processes of elastic electron scattering by endohedral $A^+@C_{60}$ positive ions.

**1. Introduction**
The idea that a phenomenological potential $U(r)$ formed by carbon atoms smeared inside a spherical layer between two concentric spheres can describe electron interaction with fullerene $C_{60}$ is a widely used approach (see, for example [1] and references therein) despite the fact that this approach is an essential simplification of the real molecular field. The inner volume of the $C_{60}$ shell is large enough to accommodate individual atoms or even small molecules. The Van der Waals forces acting between the electrically neutral encapsulated atom and the $C_{60}$ shell are too weak to distort the electronic structure of both the atom and the shell itself, and therefore these structures can be considered as independent of each other.

The situation changes as a result of, for example, photoionization of encapsulated atom [2], when in the shell center a positive charge arises. In this case the positive electric field of ion shifts the negative electron density of the $C_{60}$ shell relative to positive density of carbon ions. The positive electric charge of the latter is smeared on the surface of the sphere with the radius $R$. Here $R$ is the distance of the carbon atoms nuclei from the center of the $C_{60}$ shell. The shifting of the electron density in each elementary volume of the $C_{60}$ shell under the action of positive atomic residue $A^+$ results in creating an induced electric dipole moment of this volume. The axes of all elementary dipole moments are directed to the center of the $C_{60}$ sphere and the electric component of the shell, as a whole, is shifted to the sphere center, which leads to the monopole polarization of the shell and in its turn changes a shape of the $C_{60}$ static potential. As long as the photoelectron is inside the fullerene cavity, the distortion of the $C_{60}$ shell potential is not enough for an electron to pass through the shell. Therefore, the effects analyzed in [2] should be considered overevalued, since the authors of [2] assume that the polarization of the shell is as if the photoelectron had already leaves out of it. However, the very problem of the $C_{60}$ shell monopole polarization by a positive charge localized at its center is of interest, for example, when studying the processes of elastic electron scattering by endohedral $A^+@C_{60}$ positive ions. The present paper is devoted to the consideration of this issue.

In the next section the spatial distributions of electric charges forming the $C_{60}$ shell will be analyzed with the Poisson equation. Our attention in this section will be focused on a modified variant of the rectangular model potential proposed in [2]. According to the authors' opinion, the corrections to usual square-well potential are capable to describe monopole polarization of the $C_{60}$ shell. In section 3 some new model potentials for the $C_{60}$ shell are discussed. It will be shown that monopole polarization of the shell as a result of its collectivized electrons shifting relative to positive carcass of $C_{60}$ could be described by variation of parameters of these model potentials. Section 4 is Conclusions.

**2. Modified version of the spherical rectangular potential well [2]**
In paper [2] where, as far as we know, the monopole polarization of the $C_{60}$ shell was considered for the first time, to describe effect a modified version of the spherical rectangular potential well was supposed (the authors of [2] call this effect as "interior static polarization of the $C_{60}$ shell"). Their potential function has the following form

$$U^*(r) = \begin{cases} \dfrac{\alpha}{r_0} - \dfrac{\alpha}{r_0 + \Delta}, & \text{if } r \leq r_0; \\ -U_0 + \dfrac{\alpha}{r} - \dfrac{\alpha}{r_0 + \Delta}, & \text{if } r_0 \leq r \leq r_0 + \Delta; \\ 0, & \text{otherwise.} \end{cases} \quad (1)$$

Here $r_0$ denotes the inner radius, $\Delta$ is the thickness, and $U_0$ is the depth of the $C_{60}$ potential well; the parameter $\alpha$ is equal either to zero, $\alpha=0$, or to 1, if the monopole static polarization is entirely ignored or complete included, respectively.

Let us analyze how the distribution of electric charges formed the $C_{60}$ shell changes (as compared with the usual rectangular potential well) when the monopole static polarization in (2) is completely included. The energy of e-$C_{60}$ interaction (2) is connected with the potential of the electrostatic field $\varphi(r)$ of the $C_{60}$ shell by the relation $U^*(r) = -\varphi(r)$. Here, taking into account that the electron charge is equal to -1[*]. The Poisson equation defines the electrostatic field potential $\varphi(r)$ is

$$\Delta U^*(r) = 4\pi\rho, \quad (2)$$

where $\rho(r)$ is the density of the electric charges forming the spherically symmetric potential well (1). It is easy to see that the additional Coulomb-potential-like and constant terms in (1) do not change the mutual disposition of electric charges forming the $C_{60}$ shell. Indeed, let us apply the Laplacian $\Delta$ from the Poisson equation (2) to the additional terms in Eq.(1). For the first line we have

$$\Delta\left(\frac{\alpha}{r_0} - \frac{\alpha}{r_0+\Delta}\right) \equiv 0. \tag{3}$$

---
[*] We employ the atomic units (at. un.) throughout the paper.

For the second line one has

$$\Delta\left(\frac{\alpha}{r} - \frac{\alpha}{r_0+\Delta}\right) = -4\pi\alpha\delta(\mathbf{r}), \tag{4}$$

since the Coulomb potential $\alpha/r$ is the Green function for the Poisson equation [3]. Again, we have zero in the right side of Eq.(2) because $\mathbf{r}\neq 0$ in this line. Thus, the additional terms in the potential function (1) do not describe changes in the mutual disposition of electric charges in the $C_{60}$ shell, as well as static monopole polarization of the fullerene shell by the additional electric charge located in the center of the shell.

We will not dwell here on the results of the operator's $\Delta$ action on the potential of the usual spherical rectangular well because they were described in detail in [4-6]. We only note that in order for the interaction of an electron with the shell to be represented by a rectangular potential, we have to assume that this potential (the first term in the middle line of Eq.(1)) is created by two concentric spheres with radiuses $r = r_0$ and $r_0+\Delta$, each with a double electric layers having zero thicknesses, which is in the rigid contradiction with the experimental data about the $C_{60}$ shell construction.

It is interesting to trace the emergence and development of the concept of rectangular shell potential of the $C_{60}$ shell. In one of the first papers [7] we read: "We introduce a shell of positive rigid background charge, jellium, which is symmetrically placed with respect to the radius $R$ of the $C_{60}$ molecule"; i.e., the positive charge of the shell is uniformly filling the spherical layer between two concentric spheres. If the authors [7] suggested that the electrons of the shell are located in it, like a "raisins in a bun", then we would have the structure resembling the J.J. Thomson's "plum pudding model" of atom. But they went the other way and defined the electron density in the shell and its effective potential by solving the Poisson equation (2) with rigid positive jellium. As a result, the authors [7] obtained the square potential well that now is a widely used model for the $C_{60}$ shell without any mention about the assumptions made.

Let us pay attention to the two different types of the smearing of carbon atoms of the $C_{60}$ shell. First, the positive charge of the nuclei along with the negative charge of the electrons is smeared on the molecule volume. In this case we have for $C_{60}$ the potential well with flat bottom and onion-like molecular structure. Second, since the nuclei of carbon atoms are located at equal distances from the center of the fullerene cage their charge smears on the sphere of the radius $R$ (but not in the volume), while the negative charge is localized in a spherical layer having a thickness of the order of the diameter of the carbon atom. Such an arrangement of the positive and negative components of the fullerene shell leads to the non-flat bottom of the well and to appearance of a minimum of the function $U(r)$ at $r=R$ [5].

Since the model of a rectangular potential well corresponds to unphysical onion-like molecular structure, attempts to improve it to describe the monopole polarization of the shell are useless. In the next section we will consider more realistic potential functions $U(r)$, changing the parameters of which describes the monopole polarization of the $C_{60}$ fullerene shell.

**3. Model potentials for $C_{60}$ shell**

The following requirements are used to select a model of the potential well $U(r)$ that properly describes the $C_{60}$ shell. The potential $U(r)$ has to be attractive and an *s*-level should exist in it with the binding energy equal to $E_s$=-2.65 eV that is the experimental value of the electron affinity energy of $C_{60}^-$. The *p*-like bound state can be considered as a ground state [8] provided that the extra electron interaction with the field of electromagnetic radiation is neglected. The function $U(r)$ should be localized in a rather thin spherical layer with the thickness $\Delta$ of about few atomic units in the vicinity of the fullerene radius *R*. As shown in [5], in order to avoid the unphysical splitting of positive charge of the $C_{60}$ shell into the two concentric spheres we have to find among the different potential functions $U(r)$ a potential well with non-flat bottom. In addition, the function $U(r)$ should exponentially decrease with the radius *r* as a potential of any neutral atomic-like system.

It is evident that the number of such potentials is unlimited. Let us consider one of them, namely the cosh-bubble potential family [4]

$$U(r) = -\frac{U_{max}}{\cosh^n[\beta(r-R)]}, \quad (5)$$

that was called so in analogy with the Dirac-bubble potential [8]

$$U(r) = -U_0 \delta(r-R). \quad (6)$$

The function (5) exponentially decreases with the radius *r* and obeys all above-mentioned requirements. In further consideration we choose for simplicity in Eq.(5) the parameter $n$=1. In the middle of the maximal depth of the well (5), the thickness of the potential well $\Delta$ is connected with the parameter $\beta$ by the following relation

$$\Delta = \frac{2}{\beta}\ln(2+\sqrt{3}) = 2.633916/\beta. \quad (7)$$

The two parameters of this potential $U_{max}$ and $\Delta$ are connected in such a way that in the potential well (5) there exists an *s*-state with above-specified energy $E_s$=-2.65 eV (for details see [4, 5]).

If we apply the Poisson equation (2) in spherical coordinates to the potential function (5) we obtain the spatial electric charge distribution that produces the potential well (5). Figure 1 presents these charge distributions for potentials with thicknesses $\Delta$=1 and 2. The charge density in figure 2 is a three-layer sandwich where the middle layer represents positively charged $C^{4+}$ ions. The inner and outer layers represent negatively charged clouds of collectivized 240 electrons of $C_{60}$. The total charge of the shell (5) is equal to zero because the cosh -bubble potential (5) is a short-range potential. In the case of $\Delta$=1, about 45% of the negative charge are located in the inner electronic cloud. The rest negative charge of the $C_{60}$ shell is localized in the outer electronic cloud. In the case of $\Delta$=2, about 40% of negative charge are in the inner cloud.

Let us show that changing the left "cheek-bone" of the potential well (5) relative to the right one corresponds to transition of the part of collectivized electrons from the outer electron cloud to the inner one and *vice versa*, i.e. this changing describes the monopole polarization of the $C_{60}$ shell. Using the Heaviside step function

$$\Theta(z) = [1+\exp(z/\eta)]^{-1}, \quad (8)$$

we replace the constant thickness $\Delta$ in Eq.(5) by the following expression

$$\Delta(r) = \Delta_L + (\Delta_R - \Delta_L)\Theta(R-r). \quad (9)$$

The diffuseness parameter $\eta$ in Eq.(8) is a fixed positive number that can be as small as we wish, and which can therefore ultimately be replaced by zero. Parameters $\Delta_L$ and $\Delta_R$ in Eq.(9) are the thicknesses of the left and right "cheek-bones" of potential well (5), respectively. Applying again Eq.(2) to the potential function (5) and taking into account Eq.(9), we obtain the charge distributions in the $C_{60}$ shell for the set of $\Delta_L$ and $\Delta_R$ parameters.

The evolution of the negative charge distribution in the left and right sides of electron clouds we see in figure 2. Solid black lines in this figure are the electronic spatial distribution with no monopole polarization of the $C_{60}$ shell ($\Delta_L=\Delta_R=\Delta$). Dash-dot magenta lines are electronic charges for the maximally considered differences of the potential well thicknesses: $\Delta_L=0.7$ and $\Delta_R=1.3$ in the upper panel and $\Delta_L=1.7$ and $\Delta_R=2.3$ in the lower one. These sets of the left ($\Delta_L$) and right ($\Delta_L$) thicknesses correspond to the maximal considered shifts of electron clouds relative the positive charges of the shell. All other curves between black and magenta lines correspond to the partial monopole polarization of the $C_{60}$ shell. Comparison of the areas under curves for inner and outer clouds shows that the charge of the inner cloud smoothly increases from 45% up to 60% in the upper panel and from 40% to 50% in the lower one when the monopole shell polarization become stronger.

Let us show that variations of "cheek-bones" of any potential wells correspond to electronic clouds shifting relative to a positive carcass of the $C_{60}$ shell. The Dirac-bubble potential (6) can be considered as a limit case (for $d \to 0$) of the Lorentz-bubble potential [4]

$$U(r) = -U_0 \frac{1}{\pi} \frac{d}{(r-R)^2 + d^2} \ . \qquad (10)$$

The maximal depth of the potential well (10) at $r=R$ is $U_{max} = U_0/\pi d$. The thickness of the potential well $\Delta$ at the middle of the maximal depth is $\Delta = 2d$. With the increase in $r$ the potential (10) decreases as $r^{-2}$. Replacing in Eq. (10) the constant thickness $\Delta$ by the step function (8) and repeating our actions done before, we obtain in figure 3 the spatial electric charge distribution that produces by the potential well (10). The shifting of electronic clouds is the same as in figure 2 (upper panel).

Thus, by varying the "cheek-bones" of the potential wells, we can describe the various degrees of the monopole polarization of the $C_{60}$ shell.

## 4. Conclusions

We have analyzed with the Poisson equation the spatial distributions of the positive charge of carbon atomic nuclei shell and negative charge of electron clouds forming the electrostatic potential of the $C_{60}$ fullerene shell as a whole. We have demonstrated that the modification of formulas for the square-well potential by means of addition to it the Coulomb-potential-like terms proposed in paper [2] does not describe the monopole polarization of the shell by the electric charge $Z$ located in the center of the $C_{60}$ shell. It has been shown that, described in paper [2], changes in the photoionization cross sections of endohedral atom A@$C_{60}$, as a matter of fact, are due to introducing into the model for the $C_{60}$ shell additional arbitrary parameters but not to changes in the mutual disposition of electric charges in the $C_{60}$ shell under the action of positive charge $Z$ arising in the photoionization process. Investigated in [6], the photoionization of H atom in molecule

H@$C_{60}$ with proper account of this polarization in the frame of the cosh-potential model Eq.(5) does not change the process cross-section. For this reason it is emphatically incorrect statement given in [2] that static monopole polarization of the $C_{60}$ shell "may not be ignored in the photoionization of endohedral atoms near threshold".

The phenomenological potentials simulating the $C_{60}$ shell potential, if they are generated by a physically reasonable three-layer charge density (see figure 1), should belong to a family of potentials with a non-flat bottom. In the paper we have proposed and discussed some model potentials for the $C_{60}$ shell. We have demonstrated that the monopole polarization of the shell by an extra inner electric charge is described by the parameter variation of these model potentials. Constructed here, the model potentials for the $C_{60}$ shell can be used to describe the processes of elastic electron scattering by endohedral $A^+$@$C_{60}$ positive ions.

**Acknowledgments**
ASB is grateful for the support to the Uzbek Foundation Award OT-Ф2-046.

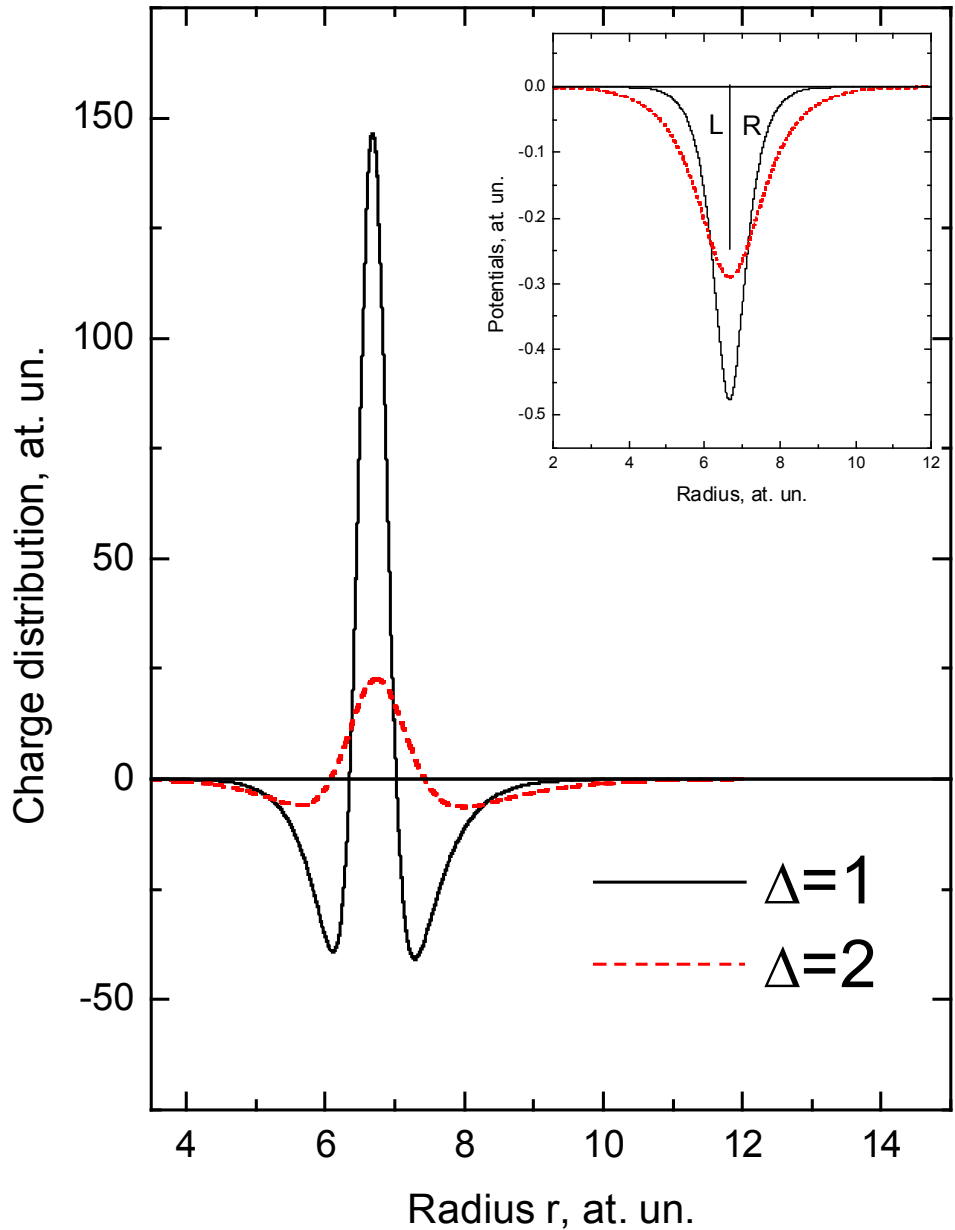

Fig. 1. Charge distribution for the potential functions Eq.(5); insert is the Cosh-bubble potential wells; letters *L* and *R* are the left and right "cheek-bones" of the potential wells Eq.(5). The parameters of wells: for thickness $\Delta=1$, depth $U_{max}=0.4762$; for thickness $\Delta=2$, depth $U_{max}=0.2898$; radius of potential wells $R=6.665$; all in at. units.

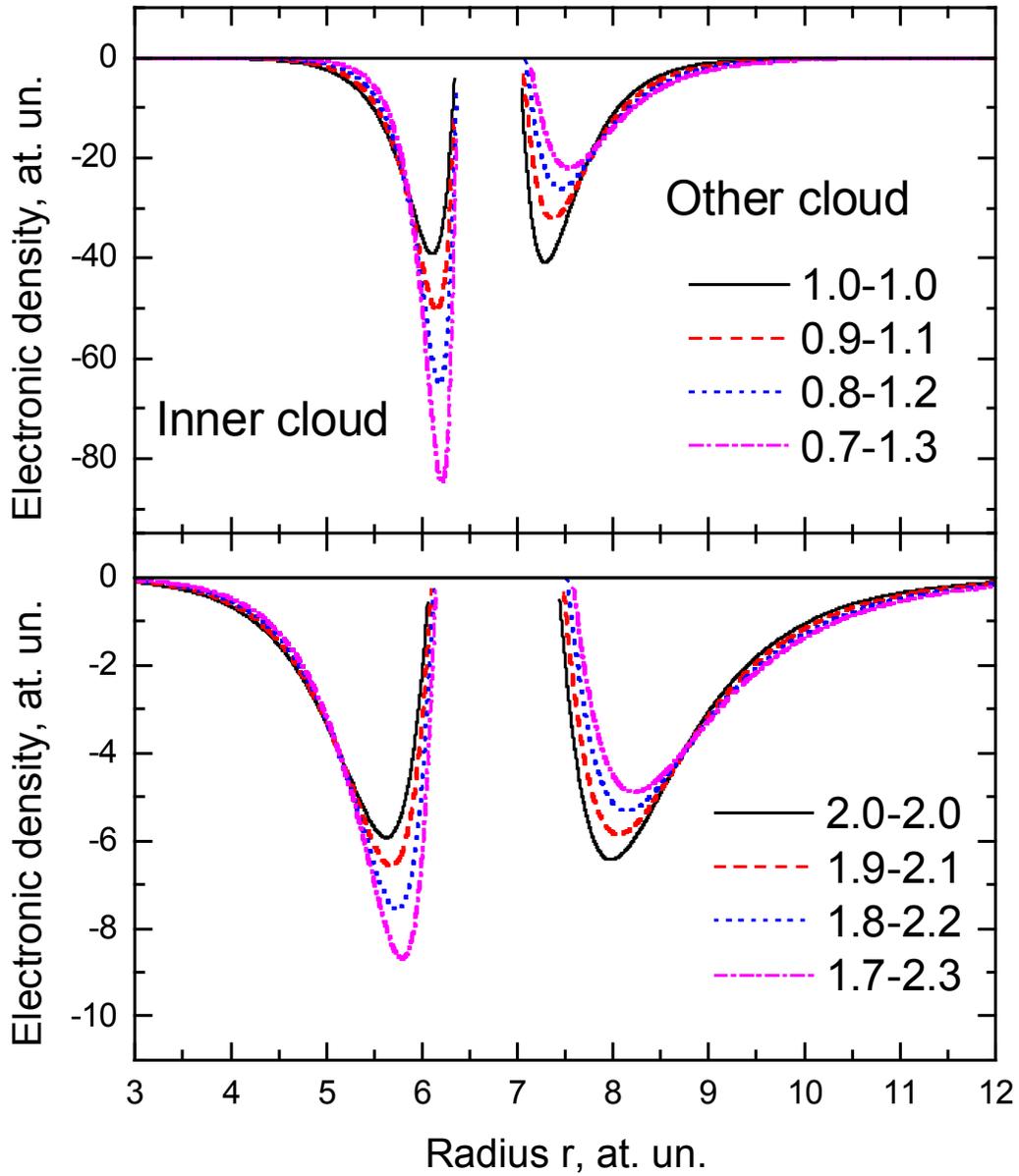

Fig. 2. Spatial distributions of negative charges in the inner and outer electronic clouds for the following combinations of the cosh-bubble potential thicknesses. In the upper panel: the line 1.0-1.0 corresponds to the combination $\Delta_L=1.0$ and $\Delta_R=1.0$; the line 0.9-1.1 to combination $\Delta_L=0.9$ and $\Delta_R=1.1$; *etc*. The same is for the lower panel. The middle part of curves (the positive charge distribution) is almost the same for all combinations of thicknesses. All parameters are in at. units.

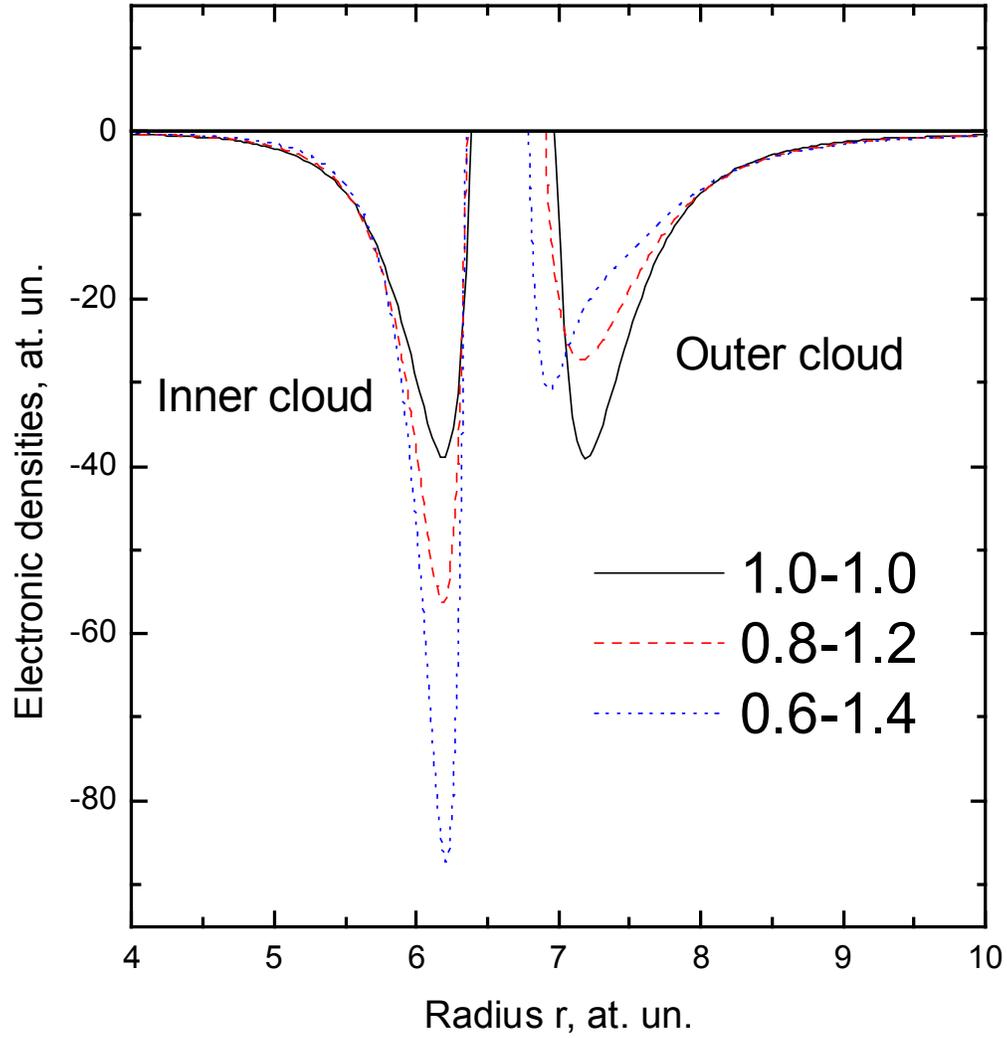

Fig. 3. Spatial distributions of negative charges in the inner and outer electronic clouds for the Lorentz-bubble potential well Eq. (10) with thicknesses: line 1.0-1.0 for the combination $\Delta_L=1.0$ and $\Delta_R=1.0$; the line 0.8-1.2 for the combination $\Delta_L=0.8$ and $\Delta_R=1.2$; *etc*. The middle part of curves (the positive charge distribution) is almost the same for all combinations of thicknesses. All parameters are in at. units.